\documentclass[usenatbib]{mnras}
\usepackage[T1]{fontenc}
\usepackage{ae,aecompl}

\usepackage{graphicx}
\usepackage{natbib}
\usepackage{amssymb}

\usepackage{hyperref}

\usepackage{etoolbox}
\makeatletter
\patchcmd\@combinedblfloats{\box\@outputbox}{\unvbox\@outputbox}{}{%
  \errmessage{\noexpand\@combinedblfloats could not be patched}%
}%
\makeatother

\newcommand{\myarcsec}{\hbox{$.\!\!^{\prime\prime}$}}
\newcommand{\myarcmin}{\hbox{$.\!\!^{\prime}$}}
\newcommand{\ks}{\hbox{$K_{\rm s}$}}
\newcommand{\myarcsecnodot}{\hbox{$\;\!\!^{\prime\prime}\;$}}

\title[MCAO imaging of distant galaxies]
      {Multi-conjugated adaptive optics imaging of distant galaxies --
        A comparison of Gemini/GSAOI and VLT/HAWK-I data}
\author[M. Schirmer et al.]{
  Mischa Schirmer$^{1}$\thanks{E-mail: mschirme@gemini.edu},
  Vincent Garrel$^{1,2}$,
  Gaetano Sivo$^{1}$,
  Eduardo Marin$^{1}$
\newauthor
  and Eleazar R. Carrasco$^{1}$
\\
$^{1}$Gemini Observatory, Casilla 603, La Serena, Chile\\
$^{2}$Max-Planck-Institut f{\"u}r extraterrestrische Physik, 85748 Garching, Germany
}

\date{Accepted XXX. Received YYY; in original form ZZZ}

\pubyear{2017}

\begin{document}
\label{firstpage}
\pagerange{\pageref{firstpage}--\pageref{lastpage}}
\maketitle

\begin{abstract}
  Multi-conjugated adaptive optics (MCAO) yield nearly diffraction-limited images at
  $2$\,$\mu$m wavelengths. Currently, GeMS/GSAOI at Gemini South is the only MCAO
  facility instrument at an 8m telescope. Using real data and for the first time, we
  investigate the gain in depth and S/N when MCAO is employed for $K_{\rm s}$-band
  observations of distant galaxies. Our analysis is based on the \textit{Frontier Fields}
  cluster MACS J0416.1$-$2403, observed with GeMS/GSAOI (near diffraction-limited) and
  compared against VLT/HAWK-I (natural seeing) data. Using galaxy number counts, we show
  that the substantially increased thermal background and lower optical throughput
  of the MCAO unit are fully compensated for by the wavefront correction, because the
  galaxy images can be measured in smaller apertures with less sky noise. We also performed
  a direct comparison of the signal-to-noise ratios (S/N) of sources detected in both data
  sets. For objects with intrinsic angular sizes corresponding to half the HAWK-I
  image seeing, the gain in S/N is 40 per cent. Even smaller objects experience a boost in
  S/N by a up to a factor of 2.5 despite our suboptimal natural guide star configuration.
  The depth of the near diffraction limited images is more difficult to
  quantify than that of seeing limited images, due to a strong dependence on the intrinsic
  source profiles. Our results emphasize the importance of cooled MCAO systems for
  \ks-band observations with future, extremely large telescopes.
\end{abstract}


\section{Introduction}{\label{intro}}
Atmospheric turbulence reduces the depth and resolution of ground-based imaging.
Multi-conjugated adaptive optics \citep[MCAO, see][]{rmv00,elr00} compensates the
wavefront distortion and reduces plate-scale dynamical distortions by using
information from several natural guide stars (NGS) as well as laser guide stars (LGS).
The corrections are sent to two or more deformable mirrors conjugated to different
altitudes. Nearly diffraction-limited images can be achieved over arcminute scales in the
near-infrared (NIR).

A MCAO demonstrator (MAD) was installed at the Very Large Telescope (VLT) for several 
runs in 2008. A survey of the results is presented by \citet{mma12}, highlighting the
success of the technology, but also a strong focus on targets with resolved stellar
populations. Results for the two extragalactic observations (COSMOS field, Chandra Deep
Field South) have not yet been published.

The completion of the Gemini Multi-Conjugate Adaptive Optics System
\citep[GeMS,][]{rnb14,nrv14} and the Gemini South Adaptive Optics Imager
\citep[GSAOI,][]{mhs04,cem12} commissioning at the end of 2012 marked the
installation of the first facility MCAO system at an 8m telescope.
Two deformable mirrors in GeMS/GSAOI are conjugated to turbulence in the ground
layer and at 9 km altitude, respectively. The system has been in full science
operation since then. Similar technology is currently tested at the Large Binocular
Telescope (LBT) with LINC-NIRVANA \citep{hrb16}. Other multiple guide star AO
systems currently in preparation, and correcting the ground layer only, are LBT/ARGOS
\citep{orb16} and VLT/GRAAL+GALACSI \citep{kma16}.

Of the 116 proposals accepted for GeMS/GSAOI in semesters 2013A through 2017A, 63 per
cent have their focus on galactic targets, and 20 per cent on nearby (i.e. well-resolved,
angular diameters $\gtrsim10^{\prime\prime}$) extragalactic sources. The remaining 17 per
cent aim at distant galaxies with angular diameters comparable to the natural seeing disk
and below. The strong focus on galactic sources is partially driven by the availability of
suitably bright ($R<15.5$\,mag) NGS asterisms, often absent at high galactic
latitudes.\footnote{An upgrade to the GeMS NGS wavefront sensors is projected
  for 2018. With a brightness limit of $R\sim18$\,mag, sky coverage will improve
  at all galactic latitudes.}

\begin{figure}
  \includegraphics[width=1.0\hsize]{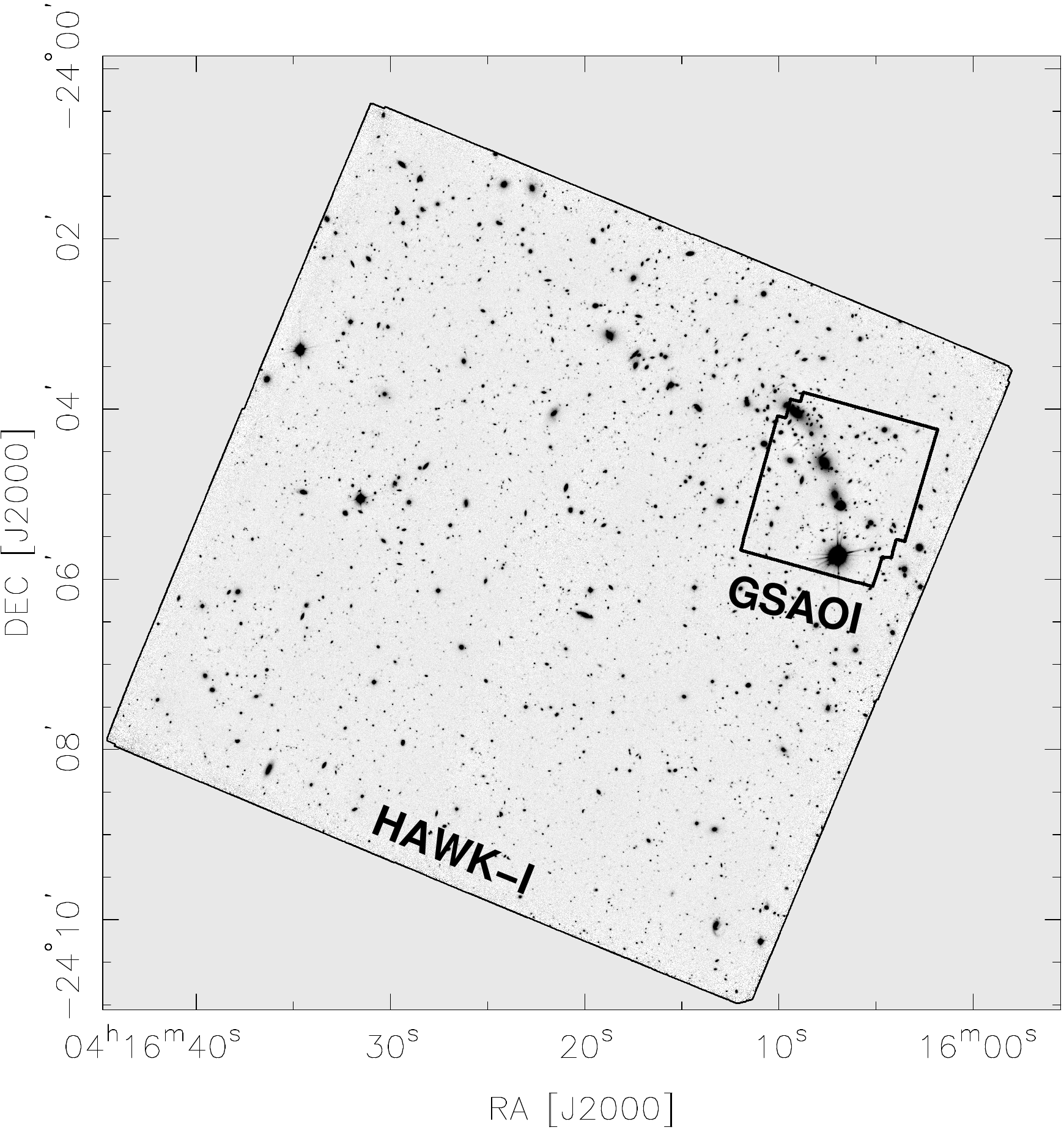}
  \caption{\label{hawki_gsaoi_field}{Sky positions of the two comparison data sets.}}
\end{figure}

For the galactic and nearby extragalactic targets, the focus with GeMS/GSAOI is mostly on 
resolved stellar populations \citep[e.g.][]{bns16} and proper motions
\citep[e.g.][]{mfm16,flz17}; in other words, the photometry and location of point
sources. The majority of these programs aims at bright targets, often against a crowded
background, with typical exposure times of $0.1-1$ hour per filter. The performance of
GeMS/GSAOI in this area is well understood \citep[e.g.][]{rnb14,nlr14,nrv14,tms15,dso16}.

Much less well investigated is the gain for the remaining 17 per cent of MCAO programs
focusing on distant galaxies \citep[e.g.][]{scp15,ssg17}. A study about the recovery
of morphological parameters of distant galaxies was done by \citet{nhh14}. It is based
on simulated images (S\'ersic profiles), convolved with a spatially dependent
model of the GeMS point spread function (PSF). An analysis based on real data is still
pending.

In this paper we perform a direct comparison between near diffraction limited
images and seeing limited images of the same field obtained with GeMS/GSAOI and
VLT/HAWK-I, respectively. We measure the gain in depth and signal-to-noise (S/N)
in the MCAO data of distant galaxies compared to the natural seeing images.

Our work is organized as follows: Section 2 summarizes the instrumental characteristics
of GSAOI and HAWK-I, the observations and data reduction. We analyze the images in
Section 3, and present our conclusions in Section 4.

\begin{figure}
  \includegraphics[width=1.0\hsize]{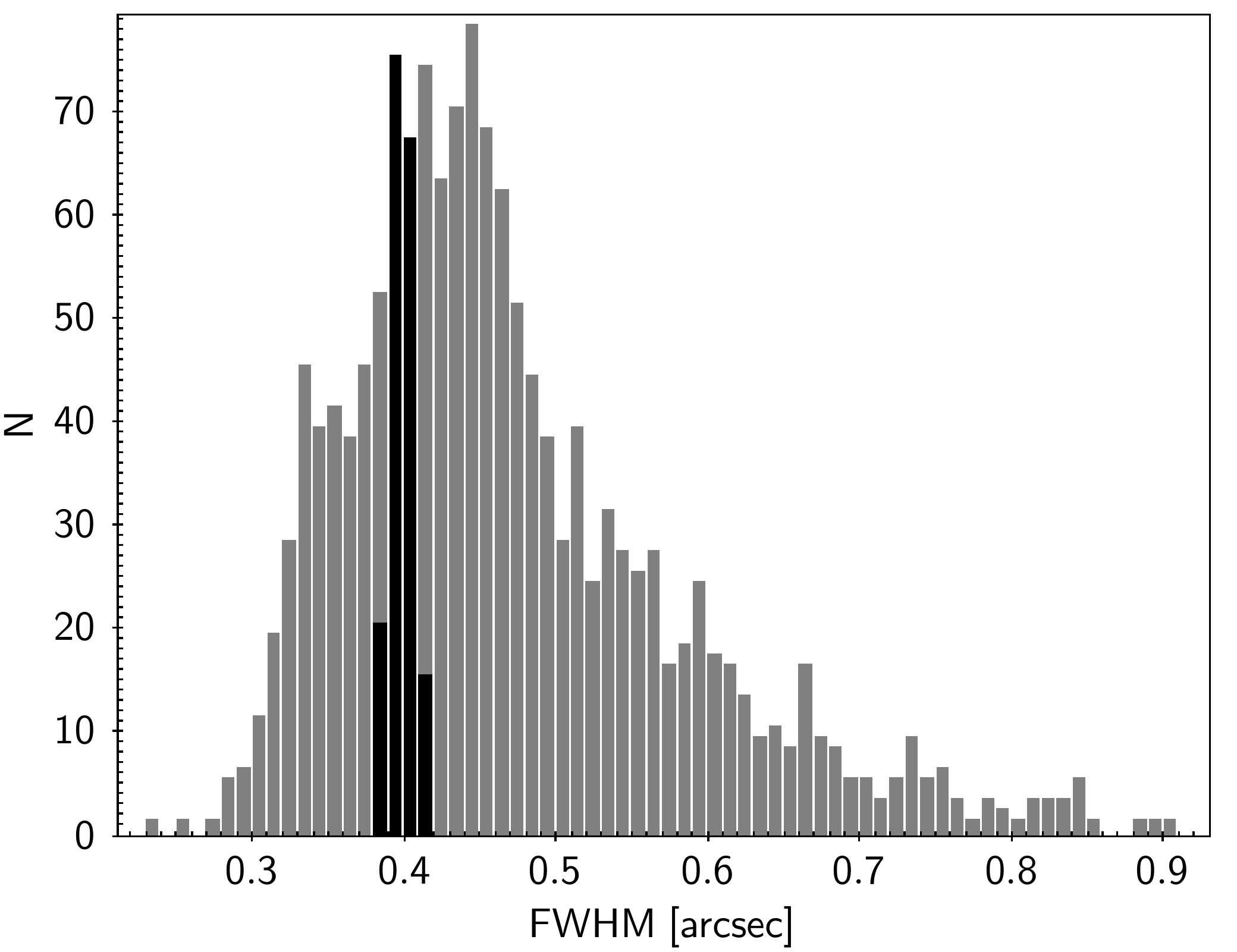}
  \caption{\label{hawki_seeingstat}{FWHM image seeing of the \textit{full} HAWK-I
      data set of MACS J0416.1$-$2403. The comparison image we use in this paper was
      constructed from a subset of 178 exposures with a seeing of
      0\myarcsec387$-$0\myarcsec413 (black histogram bars).}}
\end{figure}

\section{Observations and data reduction}
\subsection{Gemini/GSAOI and VLT/HAWK-I}
The focal plane of GSAOI \citep{mhs04,cem12} is formed by a $2\times2$ mosaic of Hawaii
2RG detectors with $2048\times2048$ pixel each; the gaps between the detectors measure 
2\farcs8$-$3\farcs0. The native pixel scale varies smoothly between 
$0\myarcsec0194-0\myarcsec0199$, and the field of view is 
$85^{\prime\prime}\times85^{\prime\prime}$. The diffraction limit of the Gemini South telescope 
is 55\,milli-arcsec (mas) in the $K_{\rm s}$-band; with optimal NGS configurations and
good laser return from the atmospheric sodium layer, a resolution of $60$\,mas has
been obtained.

Like GSAOI, the focal plane of HAWK-I \citep{pkm04,cpk06,kpc08,scv11} consists of a 
$2\times2$ mosaic of Hawaii 2RG detectors. The pixel scale is 0\myarcsec1064, a factor 
of five larger than for GSAOI, and the detector gaps are 15\myarcsecnodot wide. HAWK-I 
covers a $7\myarcmin5\times7\myarcmin5$ field of view.

The comparison in this paper is facilitated by the instruments' similar characteristics.
Both cameras use the same NIR detector type and are mounted at telescopes with $\sim8m$
apertures at similar altitudes (2700\,m for Gemini South, 2600\,m for the VLT).
The telescopes are located in Chile, where they experience similar large-scale weather
systems. Nonetheless, the local weather can differ substantially at any given time. This,
however, can be ignored for the purposes of this paper, as both data sets were taken in
very good conditions.

\subsection{Observations}
The GSAOI data were obtained under Gemini programme GS-2013B-DD-1 (PI Carrasco) on five
nights in January 2014, totaling $184\times120$\,s. Exposures with inferior corrected
seeing (one night with bad natural seeing) were rejected, resulting in a total integration
time of 15720\,s. The comparison in this work is based on the publicly released coadded
image {\tt MACS\_J0416.1-2403\_GSAOI\_0.02\_deep.fits} \citep[see table 3 in][]{scp15},
which is available online\footnote{{\tt http://www.gemini.edu/node/12254}}.

The HAWK-I data were obtained under ESO programme 092.A-0472 \citep{bml16} on 28 nights
between October 2013 and February 2014, totaling $1740\times56$\,s. Each 56\,s exposure is
constructed from 8\,s images, averaged using on-detector computations before
readout. 178 exposures were selected to construct the natural seeing image for our comparison;
details about the selection are provided below.

In both data sets, the sensitivity for all exposures is limited by thermal background.

\begin{figure}
  \includegraphics[width=1.0\hsize]{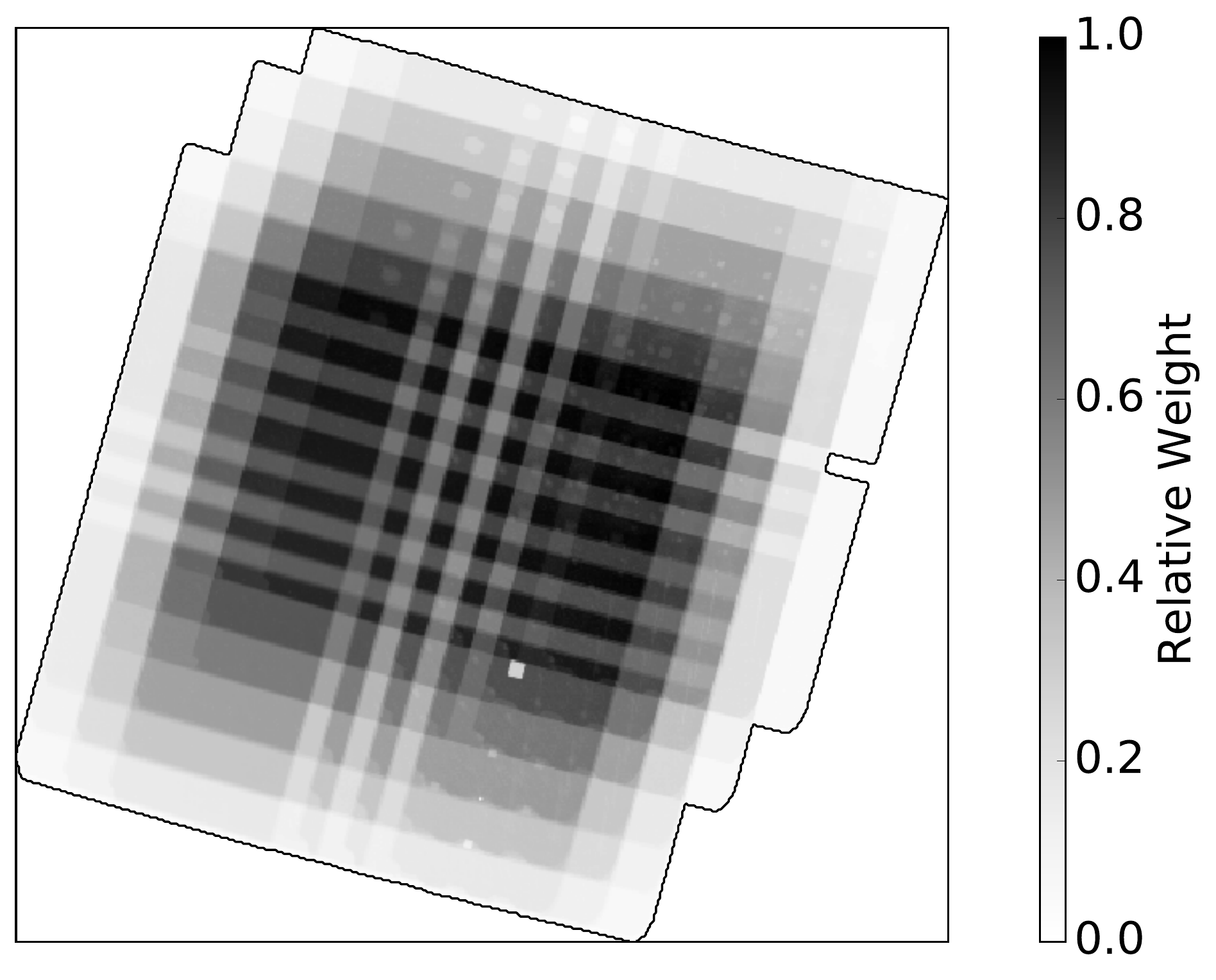}
  \caption{\label{weightmap}{Normalized weight map of the coadded GSAOI image,
      showing the uneven depth of the data. The \textit{total} integration time
      is 16\,ks, the \textit{average} exposure time is that of 10\,ks.}}
\end{figure}

\subsection{Data reduction}
All data were reduced using {\tt THELI} \citep{sch13,esd05}. Processing of the HAWK-I data 
followed NIR standard procedures including flat-fielding, dynamic two-pass background 
modeling, astrometric calibration against 2MASS \citep{scs06}, distortion correction, 
adjustment of relative photometric zeropoints to compensate for variable atmospheric 
transmission, and finally a weighted image coaddition. 

The GSAOI data were processed similarly. Full details are given in \citet{scp15}, with
emphasis on astrometric correction and background subtraction. The images were registered
against an astrometric reference grid constructed from the sources in the coadded HAWK-I
image. Thus, both data sets share the same world coordinate grid.

\subsection{Construction of the comparison images}{\label{construction}}
For our comparison to be meaningful, the two coadded images should be built from
images with stable seeing, and have the same effective exposure time and PSF sampling.
Thanks to the large number of HAWK-I images we could easily select a sub-sample
meeting these criteria:

\begin{figure}
  \includegraphics[width=1.0\hsize]{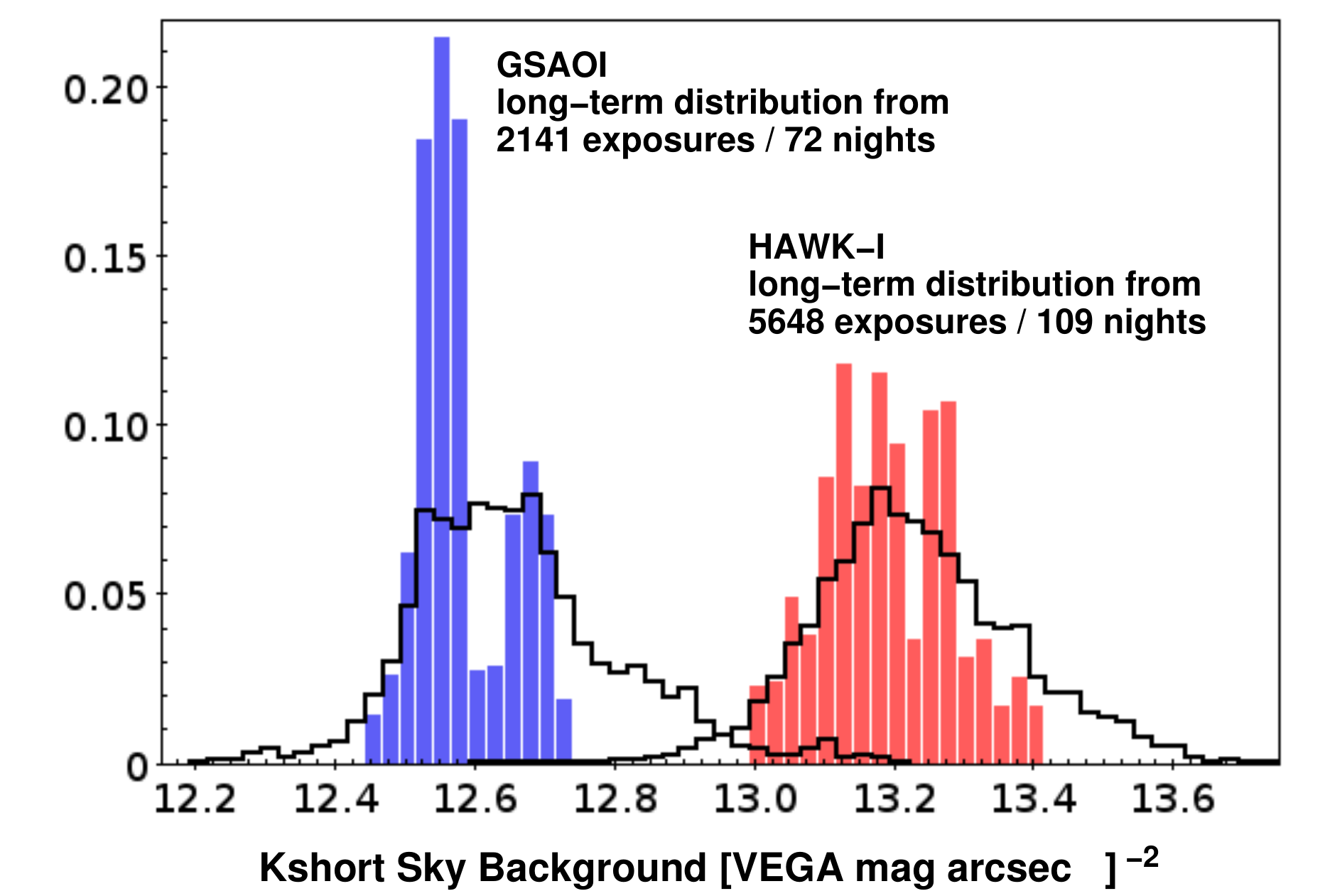}
  \caption{\label{backgroundstat}{Observed $K_{\rm s}$ thermal background for GSAOI (blue)
      and HAWK-I (red), irrespective of airmass. The additional contribution from the
      MCAO unit is evident, as the $K_{\rm s}$ sky background on Cerro Paranal (VLT)
      and Cerro Pachon (Gemini) are within 0.1 mag, otherwise.
      The blue and red histograms show the background distributions for the exposures
      that were used for the coadded images. The solid lines display the long-term
      distributions seen by both instruments between years 2013 and 2015; the data for
      our comparison were taken in representative conditions.}}
\end{figure}

The image quality of the GSAOI data is 0\myarcsec08$-$0\myarcsec10 (the natural seeing
was 0\myarcsec4$-$0\myarcsec9). From the HAWK-I data, 178 exposures were selected with a
seeing of 0\myarcsec387$-$0\myarcsec413 (Fig. \ref{hawki_seeingstat}), yielding a total
exposure time of 9968\,s.
This is lower than the total exposure time for GSAOI (16\,ks). However, the GSAOI
data are widely dithered and the detector gaps add to the non-uniformity of the
effective exposure time (Fig. \ref{weightmap}); the \textit{average} depth is that
of 10\,ks. The GSAOI data fully fit into one HAWK-I detector, and the HAWK-I exposure
time is uniform across that area (see Fig. \ref{hawki_gsaoi_field}).
Lastly, the sampling of the coadded images is also similar. There are 4.2 pixels in the full
width half maximum (FWHM) of the GSAOI image (0\myarcsec085 seeing), and 3.9 pixels
in the FWHM of the HAWK-I image (0\myarcsec41 seeing).

\begin{figure*}
  \includegraphics[width=1.0\hsize]{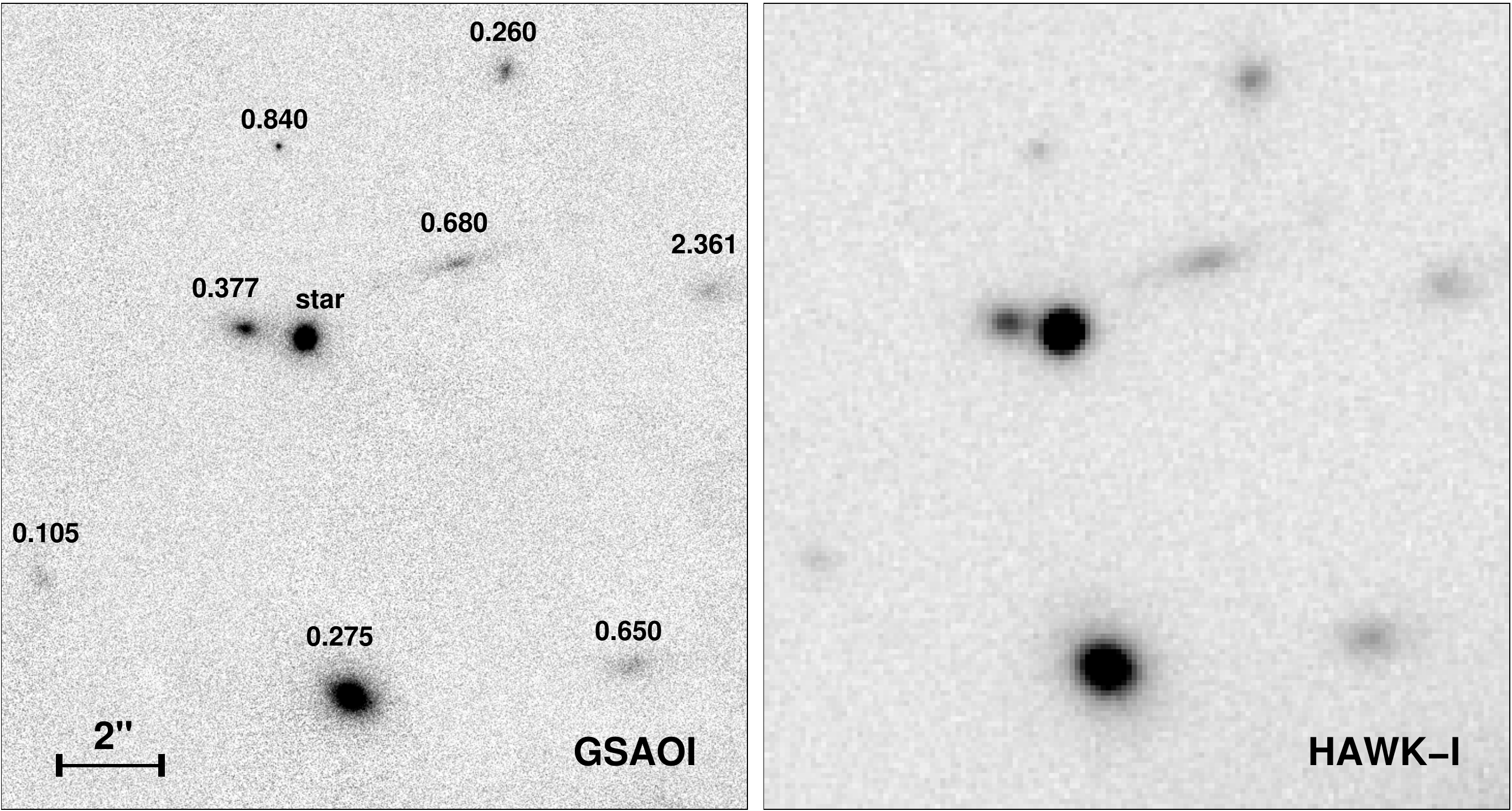}
  \caption{\label{detection_comp}{Images of distant galaxies as seen by GSAOI
      \citep[left, including photometric redshifts of][]{cam16} and HAWK-I (right,
      5 times enlarged). The detection limit for the blurred HAWK-I data is mostly
      independent of the intrinsic source profiles.}}
\end{figure*}

Note that the natural seeing of the HAWK-I exposures selected for the analysis is
narrowly clustered around 0\myarcsec4, whereas the GeMS/GSAOI data have a natural seeing
of 0\myarcsec4$-$0\myarcsec9. If the focus of the analysis was more on the actual
difference brought by GeMS, then one should select the HAWK-I images such that their
seeing distribution resembles that of the GeMS/GSAOI data. This, however, is not the
scope of our work.

\section{Analysis}{\label{analysis}}

\subsection{Reduced throughput and enhanced thermal background from MCAO}
An important difference between HAWK-I and GeMS/GSAOI is that GeMS adds one transmissive 
and eight reflective elements to the optical path. The total throughput (telescope to
detector) in $K_{\rm s}$-band for GSAOI is 31 per cent \citep{rnb14}. In comparison, for
HAWK-I we have a total \textit{instrument} throughput of $\gtrsim 50$ per cent
\citep{kpc08}.
The typical reflectivity of the VLT's aluminum-coated mirrors is 0.97 in $K_{\rm s}$-band
\citep{egs99}, and there are three such mirrors in front of HAWK-I (which is mounted
at the Nasmyth focus). Hence we estimate the total net throughput to $\sim45$ per cent.
This is significantly more than the 31 per cent for GeMS/GSAOI, and can be explained if
the eight mirrors in GeMS, and the science fold and AO fold flat mirrors (all silver
coated) have a mean reflectivity of 0.97. Nominally, silver has a reflectivity of
$0.980-0.985$ in $K_{\rm s}$-band, and perhaps it degraded since the commissioning of the
system (no measurement data available). We conclude that the MCAO system removes about
0.4 mag sensitivity mainly because of the higher number of optical surfaces.

GeMS is kept at ambient temperature, increasing the thermal background seen by GSAOI.
The average long-term $K_{\rm s}$ background for GSAOI is $12.6\pm0.2$ VEGA mag
arcsec$^{-2}$, compared to $13.2\pm0.2$ mag arcsec$^{-2}$ measured by Gemini/FLAMINGOS-2
(a NIR imaging spectrograph, which does not have the MCAO system in its
path)\footnote{The FLAMINGOS-2 statistics will be published elsewhere.}.
Our MCAO system therefore adds $\sim0.6$ mag to the sky brightness in $K_{\rm s}$,
implying 1.7 times longer integration times to reach the same S/N. This is a bit lower
than the commissioning results from \cite{cem12}, who determined the instrumental
contribution of GeMS by observing blank sky fields once through the AO system, and
once by bypassing it. They find factors of $1.5-3.5$ or, averaged over the $K_{\rm s}$
bandpass, a thermal contribution of 0.9 mag. The difference to 0.6 mag measured by our
long-term data is not surprising, given the high susceptibility of the
$K_{\rm s}$ background to ambient temperature. For example, our long-term FLAMINGOS-2
statistics shows that the $K_{\rm s}$ background at Gemini South increases by
0.05\,mag for every degree of the primary mirror temperature. Together, these are strong
arguments for cooled (yet non-cryogenic) MCAO systems such as NFIRAOS \citep{haa10}
at the Thirty Meter Telescope, and high altitude sites that provide a generally
colder climate.

Figure \ref{backgroundstat} shows the combined instrumental and sky background for
GSAOI and HAWK-I, together with their long-term distributions. The latter were
measured from archival images, showing that both data sets were obtained in typical
conditions. The background seen by HAWK-I ($13.2\pm0.2$ mag arcsec$^{-2}$) is the
same as that of FLAMINGOS-2, i.e. Cerro Paranal (VLT) and Cerro Pachon (Gemini)
share the same $K_{\rm s}$-band sky brightness.

Because of the reduced throughput and increased background, GeMS/GSAOI lose 1.0 mag
compared to a seeing limited instrument such as HAWK-I. In the rest of this paper,
we quantify how much GeMS compensates for this loss, because it concentrates the
light of distant galaxies in a smaller area with lower cumulative sky noise. This
depends on the intrinsic sizes and morphologies of these galaxies, and is best studied
with real data. Typical background galaxies with redshifts $z\gtrsim1$ have sizes of
0\myarcsec10$-$0\myarcsec25 \citep[as resolved by the Hubble Space Telescope,
  see][]{rrg96,slm12}. This is $2-4$ times smaller than the PSF of our HAWK-I image,
and hence these objects should respond particularly well to MCAO.

\subsection{Number counts}{\label{numcounts}}
One way to compare the images are number counts. The interpretation is not straight
forward though, because most distant galaxies have angular diameters smaller than the
natural seeing. Once convolved by the natural seeing PSF, the galaxy images have
approximately the same size (Fig. \ref{detection_comp}) and respond uniformly to the
detection filters. The detection limit in MCAO data, on the other hand, depends on the
sources' light profiles. A faint and compact object may easily be detected in the MCAO
data, but not in a classical image. Likewise, each pixel of an extended low surface
brightness galaxy could be pushed below the detection threshold in MCAO data (because
its light is distributed over many pixels), yet it would be detected in the classical
data with larger pixel scale.

\begin{figure}
  \includegraphics[width=1.0\hsize]{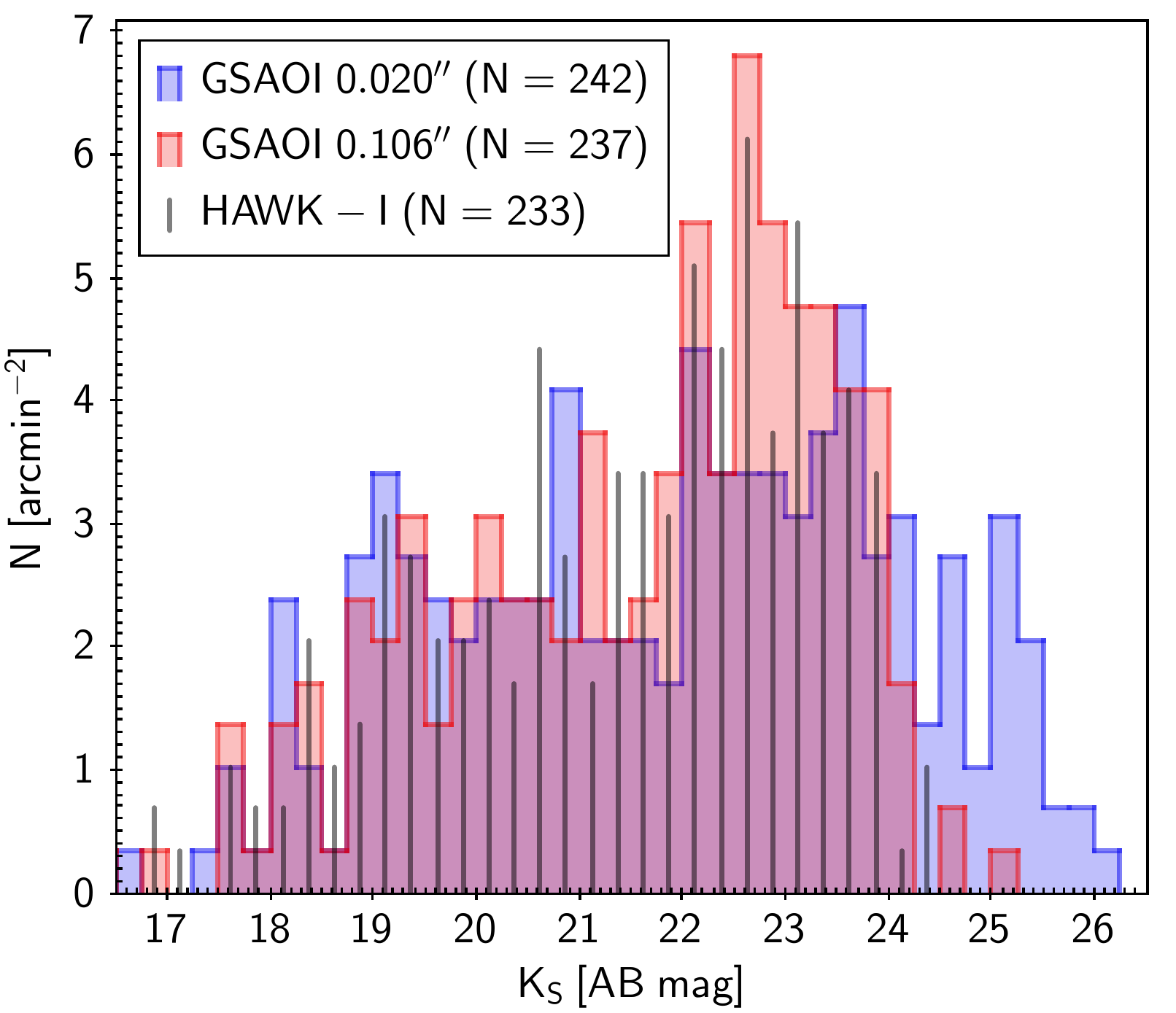}
  \caption{\label{ncounts}Number counts for the HAWK-I (vertical lines) and GSAOI
    stacks, once for the native GSAOI pixel scale (blue histogram), and once for
    the GSAOI data resampled to the HAWK-I pixel scale (red histogram). The counts
    were measured in the area common to both data sets (see Fig. \ref{hawki_gsaoi_field}).
    Down to $K_{\rm s}\sim22.5$ (completeness limit) the counts are similar, below
    that we detect more sources with GSAOI. Further details are given in the main
    text.}
\end{figure}

We probe these effects using two different coadditions of the GSAOI data, once
maintaining the native plate scale of 0\myarcsec02 pixel$^{-1}$, and once resampling to
the HAWK-I plate scale of 0\myarcsec106 pixel$^{-1}$. Object detection is done on all
three images (the two GSAOI versions, and the reference HAWK-I image) with {\tt SExtractor}
\citep{bea96}. Detection thresholds of {\tt DETECT\_THRESH} $=2.0$ and
{\tt DETECT\_MINAREA} $=4$ are applied, i.e. an object must consist of at least 4
connected pixels with S/N $\geq2$ each. Coadded weight maps are used to take local
noise properties into account. HAWK-I detections are considered only if they fall
within the GSAOI field of view (Fig. \ref{hawki_gsaoi_field}), avoiding biases by
the strong concentration of cluster galaxies.

The results are shown in Fig. \ref{ncounts}. The turn-over point (completeness limit)
is reached at $K_{\rm s}\sim22.8$ AB mag in the HAWK-I and the rebinned GSAOI image,
with approximately equal counts at brighter magnitudes. Overall, the count profile
for the HAWK-I image (vertical lines) and the rebinned GSAOI image (red histogram)
are very similar. For fainter magnitudes, more detections are being made in the GSAOI
data, in particular the one with the native pixel scale. We conclude that the wavefront
correction fully makes up for the loss in overall throughput and the enhanced background
of the MCAO.

The histograms in Fig. \ref{ncounts} reveal that the effective 5-fold rebinning lifts
pixels below the $2\sigma$ detection threshold in the unbinned image over the
threshold in the binned image. Consequently, objects in the unbinned image and with
magnitudes $24-26$ (blue histogram) become $1-1.5$ mag brighter, which is seen in
the higher number counts around $23-24$ mag (red histogram). However, the binning
also concentrates more sky noise into the measurement aperture, and thus the very
faint and smallest objects are lost compared to the unbinned image.

Our analysis verifies that the light profiles of compact galaxies affect their response
to the detection filters in near diffraction limited images. The depth of a MCAO image
is therefore no longer a simple function of magnitude as in a seeing limited image, and
the concept of ``limiting magnitude'' is less well defined. A more robust analysis of the
number counts is prohibited by the low number of detections, and would require to go
about half a magnitude deeper for this particular target area.

\begin{figure}
  \includegraphics[width=1.0\hsize]{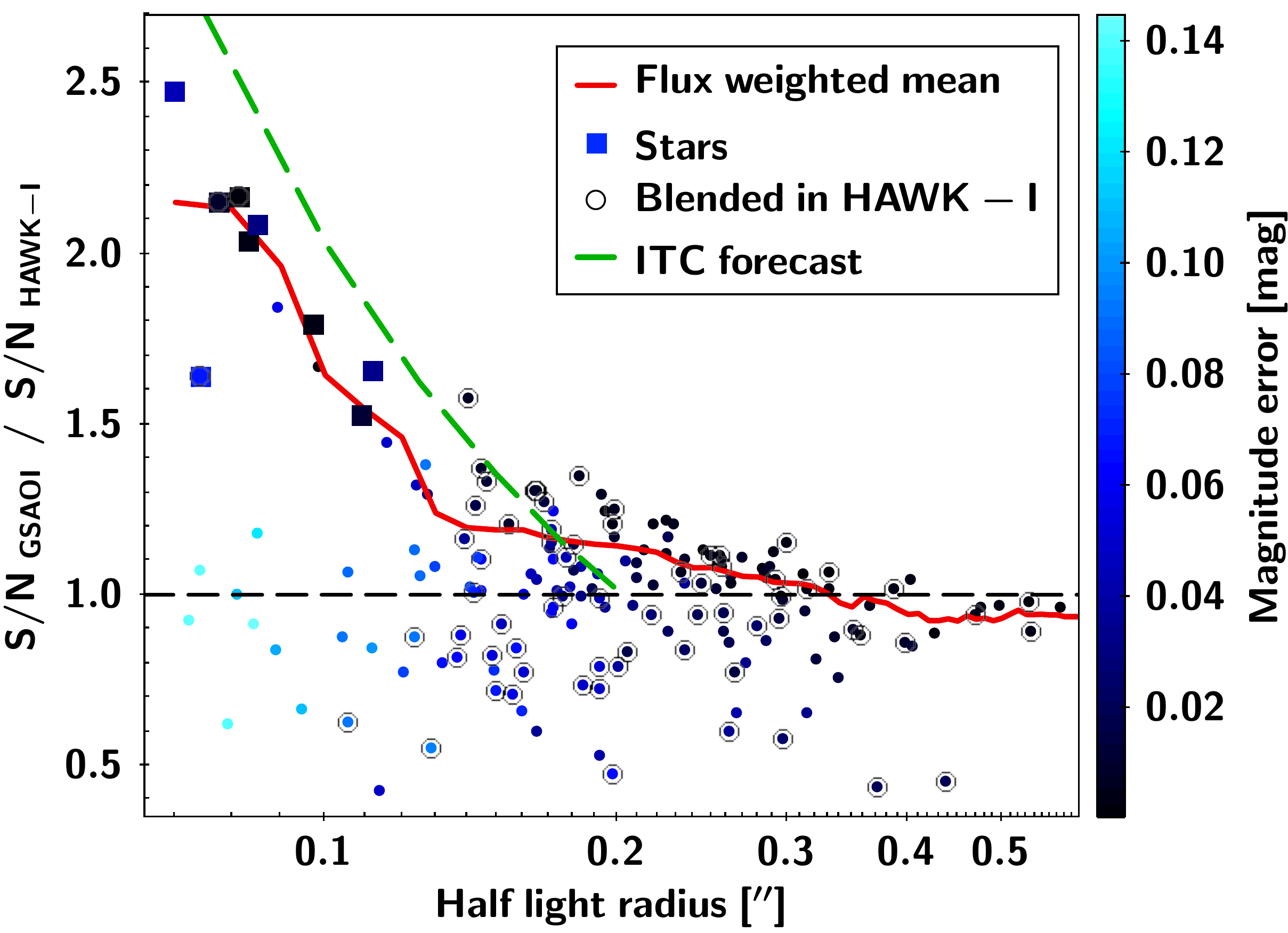}
  \caption{\label{snratio}{Ratio of S/N between GSAOI and HAWK-I as a function
      of the source half-light radius and magnitude. The HAWK-I image seeing 
      has a measured half-light radius of $0\myarcsec26$. Sources with intrinsic
      diameters of half the natural seeing disk already gain $\sim40$ per cent in S/N
      from MCAO. Even smaller sources experience a dramatic boost. Sources which
      are blended in the HAWK-I data and resolved by GSAOI are marked with open
      circles. They appear to have less gain in S/N because the combined flux
      in HAWK-I is compared with the (reduced) flux of the closest GSAOI detection.
      The dashed green line is the prediction from a comparison of the GSAOI and
      HAWK-I integration time calculators.}}
\end{figure}

\subsection{Direct S/N comparison}
A better and more direct way to quantify the gain by MCAO is to compare the S/N of
individual sources. We use the {\tt SExtractor} catalogs from Sect. \ref{numcounts}
for the unbinned GSAOI image and the HAWK-I image. Only sources that are common to
both catalogs enter the analysis.

The background noise in the GSAOI image is uneven (Sect. \ref{construction}). We must
therefore estimate the S/N an individual source would have if it did not fall into an
area with increased noise. We extract relative correction factors from the coadded
weight image (Fig. \ref{weightmap}), which reflects the noise characteristics
(${\rm weight}\propto{\rm noise}^{-2}$).

We then calculate (S/N$_{\rm GSAOI}$) / (S/N$_{\rm HAWK-I}$) for stars and galaxies, i.e.
the \textit{ratio of S/N}. It is shown as a function of the half-light radius
$r_{\rm h}$ ({\tt SExtractor}'s {\tt FLUX\_RADIUS}) and magnitude in Fig.
\ref{snratio}. We use $r_{\rm h}$ instead of FWHM because
it is less susceptive to substructure. The image seeing of HAWK-I is
$r_{\rm h}=0\myarcsec26\pm0\myarcsec01$. For sources larger than the seeing disk
there no gain is detected (as expected). Sources with intrinsic sizes of about half
the seeing value gain $\sim40$ per cent in S/N. Even smaller sources experience a
substantial boost by factors of $2.0-2.5$. Note that this performance boost is
conservative, because only a single NGS was available, resulting in 0\myarcsec072
near the guide star and 0\myarcsec122 at the largest separation \citep[see][]{scp15}.
With a better NGS configuration we would have reached $0\myarcsec06-0\myarcsec07$
over the entire field, improving the S/N further and for more sources.

Figure \ref{snratio} also reveals objects for which the MCAO correction appears
to reduce the S/N. $60-65$ per cent of those are blended in the HAWK-I data and resolved
by GSAOI. The resolved component in GSAOI that is matched with the blended
HAWK-I source has systematically lower flux. There is also a clear trend with
magnitude error: sources with higher flux are less impacted by noise, therefore
the gain by MCAO is more readily visible.

We also overplot a theoretical prediction (dashed green line in  Fig. \ref{snratio}),
using the GSAOI and HAWK-I integration time calculators (ITC). The prediction is
limited because of the inherent limitations of these tools. We use a source with
$K_{\rm s}=24$ VEGA mag and a spectrum of an elliptical galaxy mapped to redshift
1.0. The magnitude is irrelevant because we build the ratio of S/N, and so is the
type of the spectrum because the relative response of both instruments across 
the $K_{\rm s}$ band is similar.

For the HAWK-I ITC, we use a point source (always smaller than 0\myarcsec4), a
fixed (0\myarcsec4) image FWHM at an airmass of 1.2, and 1246 exposures
with 8s DITs (detector integration time, reflecting the total of 9968s). The same
is done for the GSAOI ITC, but with $83\times120$s exposures, and calculations for
a range of FWHM for an extended Gaussian source profile. The FWHM are converted to
halflight radii, and the GSAOI S/N is normalized by the S/N for HAWK-I. The prediction
breaks down for FWHM equal or larger than the HAWK-I seeing, because then our assumption
of a point source in the HAWK-I ITC is no longer valid. Overall, the prediction is
fairly good over its valid range.

\section{Summary}
We compared a unique set of observations of the same
extragalactic target, obtained with two of the most advanced NIR
imagers currently available at 8m telescopes: GeMS/GSAOI at the
Gemini South telescope, and HAWK-I at the VLT. We quantified the
performance gain delivered by MCAO over classical imaging for
distant galaxies, as a function of their intrinsic size and magnitude,
and compared with the prediction from integration time calculators.

This is the first study of this kind based on real data, and it is quite
unique. We are using a set of $K_{\rm s}$-band images of the \textit{Frontier Field}
galaxy cluster MACS J0416.1$-$2403. No other target has been observed as deep
as this one with GeMS/GSAOI (MCAO); simultaneously, it is also one of the deepest
observations that have been conducted with VLT/HAWK-I (classical imaging).
To date, this is \textit{the only} existing matching pair of sufficiently deep
observations to evaluate the performance gain of MCAO for distant extragalactic
targets with real data. Our results are as follows:

\begin{enumerate}
\item{The MCAO of GeMS causes a nominal $1.0$ mag loss in sensitivity in
  $K_{\rm s}$-band because of reduced throughput (0.4 mag; additional optical
  surfaces) and increased thermal background (0.6 mag; ambient temperature).}
\item{The wavefront correction concentrates the light of the distant
  galaxies into smaller apertures with less cumulative sky noise. Our number counts
  show that this fully compensates for the loss in throughput and enhanced sky
  background.}
\item{MCAO starts to improve the S/N of individual sources once their intrinsic
  diameters become smaller than the natural seeing. Sources half the size of the
  seeing disk already gain $\sim40$ per cent. Below this, the improvement in S/N is
  dramatic, reaching factors $2-2.5$ for the smallest and most distant galaxies.}
\item{The integration time calculators (ITCs) predict the gain in S/N with useful accuracy,
  although slightly optimistic for GeMS. The latter is at least to some extent caused
  by our imperfect NGS constellation that cannot be reflected in the ITC.}
\end{enumerate}

The performance gain we measured is a conservative lower limit. The field of
MACS J0416$-$2403 offered only one natural guide star for tip-tilt and focus
correction, preventing to reach the best image quality across the entire field of view.
With an optimal NGS constellation (3 stars in a wide triangular fashion), we would have
achieved $20-50$ per cent better image quality across the field.

We recognize the limited statistics for our analysis. The field of view of GSAOI
is comparatively small, and we do not reach as deep as the \textit{full} HAWK-I
data set. A larger survey area, and/or deeper observations with GSAOI, would have
resulted in a larger number of common sources and a clearer result for the number
counts analysis. Such data do not yet exist, but with other MCAO systems on the
horizon and the coming upgrades to GeMS (new laser and wave front sensors) this
is hopefully just a matter of time.

Lastly, we wish to emphasize the importance of future cooled (non-cryogenic)
MCAO systems with high optical throughput. They should result in a significant
performance gain, reducing the gap between ground-based observatories and the future
James Webb Space Telescope.

\section*{Acknowledgments}
The authors thank the referee, Francois Rigaut, for his comments that substantially
improved the readability and accuracy of this manuscript. Author contributions: MS
reduced and analyzed both data sets and wrote the manuscript. GS and VG maintain and
improve GeMS, and operate it at night together with EM. RC lead the proposal for the
GSAOI data.

Based on observations obtained at the Gemini Observatory, which is operated 
by the Association of Universities for Research in Astronomy, Inc., under a 
cooperative agreement with the NSF on behalf of the Gemini partnership: the 
National Science Foundation (United States), the National Research Council 
(Canada), CONICYT (Chile), Ministerio de Ciencia, Tecnolog\'{i}a e 
Innovaci\'{o}n Productiva (Argentina), and Minist\'{e}rio da Ci\^{e}ncia, 
Tecnologia e Inova\c{c}\~{a}o (Brazil). 

Based on observations made with the European Southern Observatory under Program
092.A-0472(A), Chile.

\bibliographystyle{mnras}
\bibliography{mybib}

\bsp	
\label{lastpage}
\end{document}